\documentclass[12pt]{article}

\usepackage[T2A]{fontenc}
\usepackage[utf8]{inputenc}
\usepackage[english]{babel}
\usepackage{amsmath}
\usepackage{amsfonts}
\usepackage{mathrsfs}
\usepackage{amssymb}
\usepackage{cite}

\def\theequation{\arabic{section}.\arabic{equation}}
\numberwithin{equation}{section}

\newcommand{\be}{\begin{equation}}
\newcommand{\ee}{\end{equation}}
\newcommand{\bea}{\begin{eqnarray}}
\newcommand{\eea}{\end{eqnarray}}

\newcommand{\p}[1]{(\ref{#1})}

\begin{document}

\begin{titlepage}

\vspace*{0.7cm}

\begin{center}
{\LARGE\bf Towards Lagrangian construction}

\vspace{0.4cm}

{\LARGE\bf for infinite half-integer spin field}

\vspace{1.2cm}

{\large\bf I.L.\,Buchbinder$^{1,2}$\!\!,\ \ \  S.\,Fedoruk$^3$\!\!,\ \ \
A.P.\,Isaev$^{3,4}$\!\!,\ \ \  V.A.\,Krykhtin$^{1}$}

\vspace{1.2cm}

\ $^1${\it Department of Theoretical Physics,
Tomsk State Pedagogical University, \\
634041 Tomsk, Russia}, \\
{\tt joseph@tspu.edu.ru, krykhtin@tspu.edu.ru}

\vskip 0.5cm

\ $^2${\it National Research Tomsk State  University,}\\{\em Lenin Av.\ 36, 634050 Tomsk, Russia}

\vskip 0.5cm

\ $^3${\it Bogoliubov Laboratory of Theoretical Physics,
Joint Institute for Nuclear Research, \\
141980 Dubna, Moscow Region, Russia}, \\
{\tt fedoruk@theor.jinr.ru, isaevap@theor.jinr.ru}

\vskip 0.5cm

\ $^4${\it St.Petersburg Department of Steklov Mathematical
Institute of RAS, \\ Fontanka 27, 191023 St. Petersburg, Russia}

\end{center}

\vspace{1cm}

\begin{abstract}
We formulate the conditions for the generalized fields in the space
with additional commuting Weyl spinor coordinates which define the
infinite half-integer spin representation of the four-dimensional
Poincar\' e group.
Using this formulation we develop the BRST approach and derive the
Lagrangian for the half-integer infinite spin fields.
\end{abstract}

\vspace{1cm}


\phantom{Keywords: }

\vspace{1cm}

\end{titlepage}

\setcounter{footnote}{0}
\setcounter{equation}{0}

\newpage

\section{Introduction}

Various aspects of massless infinite spin irreducible
representations of the Poincar\'{e} group
\cite{Wigner39,Wigner47,BargWigner} attract much attention last time
(see, e.g., \cite{Iv-Mack}--\cite{Naj}). Such representations
contain an infinite number of states with all possible integer or
half-integer helicities, in contrast to the usual massless
representations describing the fields of fixed helicity. New
approaches to the Lagrangian description of such fields have been
recently developed in  \cite{Mets16}, \cite{Mets17}, \cite{Zin},
\cite{HabZin}, \cite{Buchbinder:2019kuh} combining the appropriate
number of free massless fields with definite helicities. Also, we
note the BRST approach \cite{BuchKrTak} to Lagrangian construction
for bosonic massless infinite spin fields. However, many of
interesting points related to the Lagrangian formulation for
massless infinite spin fields still deserve further study. In this
paper, we consider the BRST approach to Lagrangian formulations of
the fermionic massless infinite spin fields.

As known, it is convenient to realize a field description of the
above representations in terms of space-time fields depending on the
additional coordinates, so that the expansion in these coordinates
gives an infinite number of the helicity states. Beginning with the
pioneer papers \cite{Wigner39,Wigner47,BargWigner}, the space-time
vector-like quantities are usually used as such additional
coordinates (see review \cite{BekSk}).

However, there is another possibility for the field description of
infinite spin particles, which uses the commuting Dirac or Majorana
spinors as the additional coordinates. For the first time, this type
of fields was considered in \cite{Iv-Mack}. In our recent papers
\cite{BFIR,BuchKrTak,BFI,BFI19a} (see also \cite{BuchGK}) we
constructed the new infinite (continuous) spin fields depending on
spinor additional coordinates. It was shown that such fields are
obtained as a result of a quantization of the special twistor
particle models.

Following  \cite{BFIR,BuchKrTak,BFI,BFI19a}, the infinite integer
spin representation is described by the field
\begin{equation}\label{psi-0}
\Psi(x;\xi,\bar\xi)\,,
\end{equation}
which depends on the space-time coordinates $x^m$ and additional
commuting Weyl spinor $\xi^{\alpha}$,
$\bar\xi^{\dot\alpha}=(\xi^{\alpha})^*$. The conditions that such
field describes the infinite spin representation are written in the
form\footnote{We use the notation as in the monograph \cite{BK} (see
also the Appendix\,A).}
\begin{eqnarray}
P^2 \Psi
&=& 0\,,
\label{psi-eq0}\\ [7pt]
\left(\xi^\alpha P_{\alpha\dot\beta}\bar{\xi}^{\dot\beta}\right) \Psi
&=& \mu \,\Psi\,,
\label{psi-eq1}\\ [7pt]
\left(\frac{\partial}{\partial \bar\xi^{\dot\alpha}} P^{\dot\alpha\beta}\frac{\partial}{\partial \xi^{\beta}}\right) \Psi
&=& -\mu \,\Psi\,,
\label{psi-eq2}\\ [7pt]
\left(\xi^\alpha\frac{\partial}{\partial \xi^{\alpha}}-\bar{\xi}^{\dot\alpha}\frac{\partial}{\partial \bar\xi^{\dot\alpha}} \right) \Psi
&=& 0\,,\label{psi-eq3}
\end{eqnarray}
where $P_m=-i\partial/\partial x^m$ and $\mu$ is a real dimensional
parameter. Infinite integer-spin field \p{psi-0} does not have any
external vector or spinor indices (see the detailed analysis of
helicity content of infinite integer- and half-integer-spin fields
in Appendix\,B). Further, we will call the relations
(\ref{psi-eq0})-(\ref{psi-eq3}) the basic conditions.

In works \cite{BFIR,BFI}, in contrast to the field \p{psi-0}, it was
proposed to describe the massless infinite half-integer spin
representation by the field with an external spinor Dirac index
$A=1,2,3,4$:
\begin{equation}\label{psi-12}
\Psi_A(x;\xi,\bar\xi)\,.
\end{equation}
In addition to the basic conditions \p{psi-eq0}-\p{psi-eq3} (written
for the field $\Psi_A(x;\xi,\bar\xi)$),  the field \p{psi-12} should
satisfy also the additional condition in form of Dirac equation
\begin{equation}\label{psi-eq12}
P_m(\gamma^m)_A{}^B\Psi_B=0.
\end{equation}
The Klein-Gordon equation \p{psi-eq0} is evident consequence of the
Dirac equation \p{psi-eq12}. In the representation $\gamma_m= \left(
  \begin{array}{cc}
    0 & \sigma_m \\
    \tilde\sigma_m & 0 \\
  \end{array}
\right)$ (see \p{gamma-matr}), the field  \p{psi-12} is represented
in terms of  the Weyl spinors
\begin{equation}\label{psi-half}
\Psi_A(x;\xi,\bar\xi)=
\left(
\begin{array}{c}
\Psi_\alpha(x;\xi,\bar\xi) \\ [7pt]
\bar{\Upsilon}^{\dot\alpha}(x;\xi,\bar\xi) \\
\end{array}
\right)
\,.
\end{equation}
For Weyl components of the field \p{psi-12}, the Dirac equation
\p{psi-eq12} are written as follows\footnote{In case of infinite
half-integer spin the equations \p{psi-eq0}-\p{psi-eq3},
\p{psi-eq12} are not sufficient to describe the irreducible massless
infinite spin representation. For irreducible representation we must
put additional constraints
\begin{equation}
\label{add-eqs-irred-12}
\frac{\partial}{\partial
\xi_{\gamma}}\,\Psi_\gamma(x;\xi,\bar\xi) \ =\ 0\,,\qquad
\frac{\partial}{\partial
\bar\xi^{\dot\gamma}}\,\bar{\Upsilon}^{\dot\gamma}(x;\xi,\bar\xi) \
=\ 0\,.
\end{equation}
These equations have been encoded in twistor formulation
developed in \cite{BFIR,BFI,BFI19a}. In the Appendix\,C we demonstrate this
statement. Further we derive the Lagrangian without considering
these constraints, assuming that they can be somehow taken into
account in the final result.}
\begin{equation}\label{psi-eq12W}
P^m\tilde\sigma_m^{\dot\alpha\gamma}\Psi_\gamma =0 \,,\qquad
P_m \sigma^m_{\alpha\dot\gamma}\bar{\Upsilon}^{\dot\gamma}=0 \,.
\end{equation}

Taking into account the basic conditions \p{psi-eq0}-\p{psi-eq3} and
the condition \p{psi-eq12}, it is natural to assume that there
exists the Lagrangian formulation where all the above conditions are
the consequences of the Lagrangian equations of motion. Diverse
approaches to the Lagrangian description of the infinite spin fields
with vector additional coordinate were considered in works
\cite{BekMou,SchToro13a,SchToro13b,SchToro13c,SchToro15,BekNajSe,Mets16,Mets17,Zin,Najafizadeh:2017tin,BekSk,HabZin,Metsaev18,Riv18,Metsaev18a,Buchbinder:2019kuh}.
One of the powerful general methods for studying the equations of
motions and Lagrangian formulations in higher spin theories is the
BRST construction which was applied to the massless infinite spin
field theory in refs.
\cite{Bengtsson13,Mets16,Mets17,AlkGr,Metsaev18,ACG18,Metsaev18a}.

In the recent paper \cite{BuchKrTak} the  BRST construction was used
to derive the Lagrangian for the infinite integer spin fields
\p{psi-0} with additional spinorial coordinate. This approach is
some generalization of the BRST construction which was used for
finding the Lagrangians of the free fields of different types in
flat and AdS spaces (see e.g.
\cite{Buchbinder1,Buchbinder:2004gp,Buchbinder:2005ua,Buchbinder:2006nu,Buchbinder:2006ge}
and the references therein, see also the review \cite{Tsulaia}).

In the present paper we develop the generalization of the BRST
method used in \cite{BuchKrTak} to derive the Lagrangian for the
infinite half-integer spin fields.

The paper is organized as follows. In Sect.\,2, we describe the
component structure of the space-time generalized fields \p{psi-12},
\p{psi-half}, depending on additional commuting spinor variables
$\xi^\alpha$, $\bar\xi^{\dot\alpha}$, and present the equations of
motion for these component fields. In Sect.\,3, we introduce the
extended Fock space in terms of additional bosonic creation and
annihilation operators and ghost operators. Then we construct the
Hermitian BRST charge and the corresponding equation of motion which
reproduce the conditions for the component fields.  Taking into
account this BRST charge, we derive the space-time Lagrangian for
fermionic infinite spin field. The resulting Lagrangian contains
both physical fields and auxiliary and gauge fields. In Sect.\,4, we
discuss the results and open issues. In Appendix\,A, we fix the
spinor notations used in this paper.  In Appendix\,B, we describe in
details the component decomposition of the infinite spin fields
which are considered in the paper. Appendix\,C is devoted to
construction of the solution to the equations \p{add-eqs-irred-12}
on the base of the twistor formalism and description of the
irreducible representation of the infinite half-integer spin filed.

\section{Spin-tensor representation for fermionic continuous spin field}

First of all we extend the generalized coordinate space ($x^m,
\xi^\alpha, \bar\xi^{\dot\alpha}$) by additional commuting Weyl
spinor $\zeta^\alpha$, $\bar\zeta^{\dot\alpha}=(\zeta^\alpha)^*$.
It allows us to replace the four-component spinor field \p{psi-half}
by the scalar field
\begin{equation}\label{psi-half-g}
\hat\Psi(x;\xi,\bar{\xi};\zeta,\bar\zeta) \ = \
\zeta^\alpha\Psi_\alpha(x;\xi,\bar{\xi}) \ + \ \bar{\zeta}^{\dot\alpha}\bar{\Upsilon}_{\dot\alpha}(x;\xi,\bar{\xi})\,.
\end{equation}
In this case the conditions \p{psi-eq0}-\p{psi-eq3} for the field
$\hat\Psi(x;\xi,\bar{\xi};\zeta,\bar\zeta)$ are rewritten in the
form
\begin{equation}
\label{psi-eq3-a} \left(\xi^\alpha
P_{\alpha\dot\beta}\bar{\xi}^{\dot\beta}-\mu\right) \hat\Psi
=0\,,\qquad \left(\frac{\partial}{\partial \bar\xi^{\dot\alpha}}
P^{\dot\alpha\beta}\frac{\partial}{\partial \xi^{\beta}}+\mu\right)
\hat\Psi =0\,,\qquad \left(\xi^\alpha\frac{\partial}{\partial
\xi^{\alpha}}-\bar{\xi}^{\dot\alpha}\frac{\partial}{\partial
\bar\xi^{\dot\alpha}} \right) \hat\Psi = 0,
\end{equation}
whereas the massless  Dirac equations \p{psi-eq12W} look like
\begin{equation}\label{psi-eq12W-a}
P^{\dot\alpha\gamma}\frac{\partial}{\partial \zeta^{\gamma}}\, \hat\Psi
=0\,,\qquad \frac{\partial}{\partial \bar\zeta^{\dot\gamma}}\, P^{\dot\gamma\alpha}\, \hat\Psi
=0 \,.
\end{equation}
The equations \p{psi-eq12W-a} imply the equation
\begin{equation}\label{psi-eq12W-KG}
P^{2}\, \hat\Psi =0 \,.
\end{equation}

In this representation for the field
$\hat\Psi(x;\xi,\bar{\xi};\zeta,\bar\zeta)$, the Hermitian angular
momentum operator $M_{mn}$ is written as follows
\begin{equation}\label{M}
M_{mn}=\sigma_{mn}^{\alpha\beta}M_{\alpha\beta}-\tilde{\sigma}_{mn}^{\dot\alpha\dot\beta}\bar M_{\dot\alpha\dot\beta}\,,
\end{equation}
where
\begin{equation}\label{M-sp}
M_{\alpha\beta} = i\xi_{(\alpha}\frac{\partial}{\partial\xi^{\beta)}}+i\zeta_{(\alpha}\frac{\partial}{\partial\zeta^{\beta)}}\,,
\qquad
\bar{M}_{\dot\alpha\dot\beta} =
i\bar{\xi}_{(\dot\alpha}\frac{\partial}{\partial{\bar\xi}^{\dot\beta)}}
+i\bar{\zeta}_{(\dot\alpha}\frac{\partial}{\partial{\bar\zeta}^{\dot\beta)}}\,.
\end{equation}
One can prove that the operator (\ref{M}) satisfies the standard
commutation relations for the Lorentz group generators.

In spinor notation, the Pauli-Lubanski pseudovector
$W_m=\frac12\,\varepsilon_{mnkl}M^{nk}P^l$ takes the form
\begin{equation}\label{W-v}
W_{\alpha\dot\alpha}=iM_{\alpha\beta}P^{\beta}_{\dot\alpha} - i\bar{M}_{\dot\alpha\dot\beta}P^{\dot\beta}_{\alpha}\,,
\end{equation}
and the second Casimir operator $W^2$ is
\begin{equation}\label{W-2}
W^2 = M^{\alpha\beta}\,\bar{M}^{\dot\alpha\dot\beta}\,P_{\alpha\dot\alpha}P_{\beta\dot\beta}
-\frac{1}{2}\left(M_{\alpha\beta}M^{\alpha\beta}+\bar{M}_{\dot\alpha\dot\beta}\bar{M}^{\dot\alpha\dot\beta}\right)P^2
\end{equation}

Using the expression \eqref{M-sp} for the angular momentum operator,
we obtain one of the possible forms for the second Casimir operator
\begin{eqnarray} \label{W2-ex}
W^2
&=&
-(\xi P\bar{\xi})\left(\frac{\partial}{\partial \bar\xi^{\dot\beta}} P^{\dot\beta\beta}\frac{\partial}{\partial \xi^{\beta}}\right)
-(\zeta P\bar{\zeta})\left(\frac{\partial}{\partial \bar\zeta^{\dot\beta}} P^{\dot\beta\beta}\frac{\partial}{\partial \zeta^{\beta}}\right)
\\ [7pt]
&& -(\xi P\bar{\zeta})\left(\frac{\partial}{\partial \bar\zeta^{\dot\beta}} P^{\dot\beta\beta}\frac{\partial}{\partial \xi^{\beta}}\right)
-(\zeta P\bar{\xi})\left(\frac{\partial}{\partial \bar\xi^{\dot\beta}} P^{\dot\beta\beta}\frac{\partial}{\partial \zeta^{\beta}}\right)
+\, \mathcal{D}\cdot P^2 \,, \nonumber
\end{eqnarray}
where the operator $\mathcal{D}$ in the last line is
$$
\mathcal{D}:=-\frac{1}{2}\left(M_{\alpha\beta}M^{\alpha\beta}+\bar{M}_{\dot\alpha\dot\beta}\bar{M}^{\dot\alpha\dot\beta}\right)
-\frac{1}{2}\left(\xi^{\alpha}\frac{\partial}{\partial \xi^{\alpha}}+\zeta^{\alpha}\frac{\partial}{\partial \zeta^{\alpha}}\right)
\left(\bar\xi^{\dot\beta}\frac{\partial}{\partial \bar\xi^{\dot\beta}}+\bar\zeta^{\dot\beta}\frac{\partial}{\partial \bar\zeta^{\dot\beta}}\right)
$$
and $(\xi P\bar{\xi}):=\xi^\alpha P_{\alpha\dot\beta}\bar{\xi}^{\dot\beta}$,
$(\xi P\bar{\zeta}):=\xi^\alpha P_{\alpha\dot\beta}\bar{\zeta}^{\dot\beta}$, \textit{etc}.

Due to the first two equations in \p{psi-eq3-a} and the equation
\p{psi-eq12W-a}, the Casimir operators of the Poincare group $P^2$
and $W^2$, defined in \p{W2-ex}, act on the field \p{psi-half-g} as
following
\begin{equation} \label{CSConditions}
P^2\hat\Psi=0\,, \qquad  W^2\hat\Psi=\mu^2\hat\Psi\,.
\end{equation}
Hence, the field \p{psi-half-g} and therefore the fields
\p{psi-half} describe the infinite half-integer spin representation.
The homogeneity operator ($U(1)$-charge), given by the last equation
in \p{psi-eq3-a}, commutes with all Poincar\'{e} generators and is
the superselection operator.

Next, we solve the first equation in \p{psi-eq3-a} (or the equations \p{psi-eq1}) as
\begin{equation} \label{ISF-exp}
\Psi_\alpha=\delta\left(\xi P\bar{\xi}-\mu\right)\Phi_\alpha(x,\xi,\bar\xi)\,,\qquad
\bar\Upsilon^{\dot\alpha}=\delta\left(\xi P\bar{\xi}-
\mu\right)\bar{\mathcal{X}}^{\,\dot\alpha}(x,\xi,\bar\xi)\,.
\end{equation}
After that, the second equation in \p{psi-eq3-a} (or the equations
\p{psi-eq2}) and the equations \p{psi-eq12W-a} (or the equations
\p{psi-eq12W}) for the fields $\Phi_\alpha$,
$\bar{\mathcal{X}}^{\dot\alpha}$ take the form
\begin{eqnarray}
&
{\displaystyle \left(\frac{\partial}{\partial \bar\xi^{\dot\alpha}} P^{\dot\alpha\beta}\frac{\partial}{\partial \xi^{\beta}}+\mu\right)\Phi_\gamma=0\,,}\qquad
&
P^{\dot\alpha\gamma}\Phi_\gamma=0\,,
\label{cond-1}\\ [7pt]
&
{\displaystyle \left(\frac{\partial}{\partial \bar\xi^{\dot\alpha}} P^{\dot\alpha\beta}\frac{\partial}{\partial \xi^{\beta}}+\mu\right)\bar{\mathcal{X}}^{\,\dot\gamma}=0\,,}\qquad
&
P_{\alpha\dot\gamma}\bar{\mathcal{X}}^{\,\dot\gamma}=0\,.
\label{cond-2}
\end{eqnarray}
We consider the solution of last equation in \p{psi-eq3-a} (or the equations \p{psi-eq3})  for the fields
$\Phi_\alpha$, $\bar{\mathcal{X}}^{\dot\alpha}$
\begin{equation}
\left(\xi^\alpha\frac{\partial}{\partial \xi^{\alpha}}-\bar{\xi}^{\dot\alpha}\frac{\partial}{\partial \bar\xi^{\dot\alpha}} \right) \Phi_\gamma
= 0\,,\qquad
\left(\xi^\alpha\frac{\partial}{\partial \xi^{\alpha}}-\bar{\xi}^{\dot\alpha}\frac{\partial}{\partial \bar\xi^{\dot\alpha}} \right) \bar{\mathcal{X}}^{\,\dot\gamma} = 0 \label{phi-eq3-a}
\end{equation}
in form of power expansion in $\xi$ and $\bar{\xi}$:
\begin{eqnarray}
\Phi_\gamma(x,\xi,\bar{\xi}) &=&
\sum_{s=0}^\infty
\frac{1}{s!}\;\varphi_{\gamma\,\alpha_1\ldots \alpha_s\dot\beta_1\ldots\dot\beta_s}(x)\;
\xi^{\alpha_1}\ldots\xi^{\alpha_s}\;\bar{\xi}^{\dot\beta_1}\ldots\bar{\xi}^{\dot\beta_s} \,, \label{Phi-expan}
\\
\bar{\mathcal{X}}^{\,\dot\gamma}(x,\xi,\bar{\xi}) &=&
\sum_{s=0}^\infty
\frac{1}{s!}\;\bar{\chi}^{\dot\gamma}{}_{\alpha_1\ldots \alpha_s\dot\beta_1\ldots\dot\beta_s}(x)\;
\xi^{\alpha_1}\ldots\xi^{\alpha_s}\;\bar{\xi}^{\dot\beta_1}\ldots\bar{\xi}^{\dot\beta_s} \,. \label{bPhi-expan}
\end{eqnarray}
The component fields $\varphi_{\gamma\,\alpha_1\ldots
\alpha_s\dot\beta_1\ldots\dot\beta_s}(x)$ and
$\bar{\chi}^{\dot\gamma}{}_{\alpha_1\ldots
\alpha_s\dot\beta_1\ldots\dot\beta_s}(x)$ are symmetric with respect
$\alpha$ and separately $\dot\beta$ indices. In this reason we use
the shorted notations for them:
\begin{equation} \label{short-notat}
\varphi_{\gamma\,\alpha(s)\dot\beta(s)}:=\varphi_{\gamma\,\alpha_1\ldots \alpha_s\dot\beta_1\ldots\dot\beta_s}\,,\qquad
\bar{\chi}^{\dot\gamma}{}_{\alpha(s)\dot\beta(s)}:=\bar{\chi}^{\dot\gamma}{}_{\alpha_1\ldots \alpha_s\dot\beta_1\ldots\dot\beta_s}\,.
\end{equation}

The equations \p{cond-1}, \p{cond-2} lead to the following equations for the component fields:
\begin{eqnarray}
&
\partial^{\dot\delta\gamma}\varphi_{\gamma\,\alpha(s)\dot\beta(s)}(x)=0\,,\qquad
&
is\,\partial^{\dot\beta_s\alpha_s}\varphi_{\gamma\,\alpha(s)\dot\beta(s)}=\mu\varphi_{\gamma\,\alpha(s-1)\dot\beta(s-1)}\,,
\label{C1+}\\ [7pt]
&
\partial_{\delta\dot\gamma}\bar{\chi}^{\dot\gamma}{}_{\alpha(s)\dot\beta(s)}(x)=0\,,\qquad
&
is\,\partial^{\dot\beta_s\alpha_s}\bar{\chi}^{\dot\gamma}{}_{\alpha(s)\dot\beta(s)}=
\mu\bar{\chi}^{\dot\gamma}{}_{\alpha(s-1)\dot\beta(s-1)}\,,
\label{C2+}
\end{eqnarray}
where we have used
$P_m=-i \partial/\partial x^m$, $\partial^{\dot\beta\alpha}=(\tilde{\sigma}^{m})^{\dot\beta\alpha}\,iP_m$.

It is convenient to combine the fields
$\varphi_{\gamma\,\alpha(s)\dot\beta(s)}$ and
$\bar{\chi}^{\dot\gamma}{}_{\alpha(s)\dot\beta(s)}$ in a single four
component object $\varphi_{C\,a(s)\dot{b}(s)}$ of the form
\begin{equation}\label{4phi}
\varphi_{C\,\alpha(s)\dot\beta(s)}=
\left(
\begin{array}{c}
\varphi_{\gamma\,\alpha(s)\dot\beta(s)} \\ [6pt]
\bar{\chi}^{\dot\gamma}{}_{\alpha(s)\dot\beta(s)} \\
\end{array}
\right)
,
\end{equation}
and define the Dirac adjoint (with respect indices ${}_C=({}_{\gamma},{}^{\dot\gamma})$ and ${}^C=({}^{\gamma},{}_{\dot\gamma})$) field as
\begin{equation}
\label{4phid}
\bar{\varphi}^C{}_{\beta(s)\dot\alpha(s)} =
\left(\chi^\gamma{}_{\beta(s)\dot\alpha(s)},
\bar{\varphi}_{\dot\gamma}{}_{\,\beta(s)\dot\alpha(s)}\right) .
\end{equation}
Hereafter, we will often omit the index ${}_C$ in four component field $\varphi_{C\,\alpha(s)\dot\beta(s)}$ \eqref{4phi}.

Similarly
 to the expression \p{psi-half-g} for the fields $\Psi_\alpha$, $\bar{\Upsilon}_{\dot\alpha}$, $\hat\Psi$,
we construct the expresion
\begin{equation}\label{psi-half-g-r}
\hat\Phi(x;\xi,\bar{\xi};\zeta,\bar\zeta)=
\zeta^\alpha\Phi_\alpha(x;\xi,\bar{\xi})+\bar{\zeta}^{\dot\alpha}\bar{\mathcal{X}}_{\,\dot\alpha}(x;\xi,\bar{\xi})
\end{equation}
for the fields $\Phi_\alpha$ and $\bar{\mathcal{X}}_{\,\dot\alpha}$
defined by \p{Phi-expan} and \p{bPhi-expan}.

In the next section we will derive the Lagrangian formulation for
the fields under consideration using the BRST approach
in terms of the four component field \eqref{4phi}.

\section{BRST Lagrangian construction}

\subsection{Generalized Fock space}

In the previous section we have used the spinor variables
$\xi^\alpha$, $\bar\xi^{\dot\alpha}$ and the corresponding momenta
given by the derivatives with respect to $\xi$, $\bar\xi$. For this
reason, we introduce the operators
\begin{equation}
\label{a-oper}
a_\alpha\,,\quad b^\alpha\,,
\end{equation}
which satisfy the algebra
\begin{equation}
\label{a-com}
[a_\alpha,b^\beta]=\delta_\alpha^\beta\,.
\end{equation}
Hermitian conjugation yield the operators
\begin{equation}
\label{c-oper}
\bar{a}_{\dot\alpha}=(a_\alpha)^\dagger\,,\quad
\bar{b}^{\dot\alpha}= (b^\alpha)^\dagger\,,
\end{equation}
with the commutation relation
\begin{equation}
\label{c-com}
[\bar{b}^{\dot\alpha},\bar{a}_{\dot\beta}]=\delta^{\dot\alpha}_{\dot\beta}\,.
\end{equation}
Below, similar to \p{short-notat}, we use the notations
\begin{equation}
\label{not-oper}
a_{\alpha(s)}:=a_{\alpha_1}\ldots a_{\alpha_s}\,,\quad \bar a_{\dot\alpha(s)}:=\bar a_{\dot\alpha_1}\ldots \bar a_{\dot\alpha_s}\,,\quad
b^{\alpha(s)}:=b^{\alpha_1}\ldots b^{\alpha_s}\,,\quad \bar b^{\dot\alpha(s)}:=\bar b^{\dot\alpha_1}\ldots \bar b^{\dot\alpha_s}\,.
\end{equation}

Following \p{a-com} and \p{c-com} we consider the operators
$a_\alpha$ and  $\bar{b}^{\dot\alpha}$ as annihilation operators and
define the "vacuum" state
\begin{equation}\label{vac}
|0\rangle\,, \qquad \langle0|=(|0\rangle)^\dagger\,,\qquad\langle0|0\rangle=1
\end{equation}
by the relations
\begin{equation}\label{caoperators}
a_\alpha|0\rangle=\bar{b}^{\dot\alpha}|0\rangle=0\,,
\qquad\langle0|\bar{a}_{\dot\alpha}=\langle0|b^\alpha=0
\,.
\end{equation}

Let us define the auxiliary Fock space with the vectors of the form
\begin{equation} \label{GFState}
|\varphi_C\rangle=
\sum_{s=0}^{\infty}|\varphi_{C\,,s}\rangle \,,\qquad
|\varphi_{C,s}\rangle:=\frac{1}{s!}\,\varphi_{C\,\alpha(s)}{}^{\dot\beta(s)}(x)\ b^{\alpha(s)}\,\bar{a}_{\dot\beta(s)}|0\rangle \,.
\end{equation}
Then the  conjugate vector to (\ref{GFState}) is written as follows
\begin{equation} \label{bGFState}
\langle\bar{\varphi}^C|=
\sum_{s=0}^{\infty}\langle\bar{\varphi}^C_s|\,,\qquad
\langle\bar{\varphi}^C_s|:= \frac{1}{s!}\,\langle 0|\,\bar{b}^{\dot\alpha(s)}\,a_{\beta(s)}\ \bar{\varphi}^{C\,\beta(s)}{}_{\dot\alpha(s)}(x).
\end{equation}
These expansions \p{GFState} and \p{bGFState} contain an equal
number of operators with undotted and dotted indices, like the
expressions \p{Phi-expan} and \p{bPhi-expan}. It is natural to
consider that the creation and annihilation operators are realized
in the space of the vectors
\p{GFState} and \p{bGFState} with external Dirac index $C=1,2,3,4$.

Let us introduce the following $(4\times 4)$ matrix operators
\begin{eqnarray} \label{T0}
(T_0)_C{}^D &:=& i \not\!\partial_C{}^D\,,\\ [6pt]
(L_0)_C{}^D &:=& l_0\,\delta_C{}^D\,,\label{L0}\\ [6pt]
(L_1)_C{}^D &:=& (l_1-\mu)\,\delta_C{}^D\,,\label{L1}\\ [6pt]
(L^+_1)_C{}^D &:=& (l_1^+-\mu)\,\delta_C{}^D\,. \label{L1+}
\end{eqnarray}
where
\begin{equation} \label{l-def}
l_0:=\partial^2=\Box\,,\qquad  l_1:=i\,a^\alpha\bar{b}^{\dot\beta}\partial_{\alpha\dot\beta}\,,\qquad
l_1^+:=i\,b^\alpha\bar{a}^{\dot\beta}\partial_{\alpha\dot\beta}\,.
\end{equation}
In what follows we will often omit the four-component indices $C,D$
in the operators \eqref{T0}--\eqref{L1+} also. The nonzero
(anti)commutators of the above matrix operators are
\begin{equation}\label{algebra}
[L_1^+,L_1]=(N+\bar{N}+2)\,L_0\,,
\qquad
\{T_0,T_0\}=2L_0 \,,
\end{equation}
where $\{L,T\} \equiv L \cdot T + T \cdot L$, and
\begin{equation}
N=b^\alpha a_\alpha\,,\qquad
\bar{N}=\bar{a}_{\dot\alpha}\bar{b}^{\dot\alpha}\,.
\end{equation}
All other (anti)commutators among the operators
\eqref{T0}-\eqref{L1+} vanish.

One can show that the vector $|\varphi_C\rangle$ \eqref{GFState}
reproduces the fermionic infinite spin equations (\ref{C1+}),
(\ref{C2+}) if the constraints
\begin{equation}\label{l0l1}
(T_0)_C{}^D|\varphi_D\rangle=0\,, \qquad
(L_1)_C{}^D|\varphi_D\rangle=0
\end{equation}
on the vector $|\varphi_C\rangle$ are imposed. Further, by using
BRST procedure, we will construct the Lagrangian, which reproduces
the conditions \p{l0l1} as the equations of motion.

\subsection{BRST charge}

Let us consider the operators $F_a$ = ($T_0$, $L_0$, $L_1$,
$L_1^+$), defined in \p{T0}-\p{L1+}, as operators of the constraints
of some yet unknown Lagrangian theory. Since these operators form
closed (super)algebra $[F_a,F_b\}=f_{ab}{}^c F_c$ \eqref{algebra} we
can build BRST charge in  a standard way as
\begin{equation}\label{Q-gen}
Q=c^aF_a+\frac12\,(-1)^{n_a+n_b}f_{ab}{}^c c^ac^b\mathcal{P}_c\,,\qquad Q^2=0\,,
\end{equation}
where $c^a$ and $\mathcal{P}_a$ are the ghosts and their momenta and $n_a=0$ or $1$ is the parity of
the operator $F_a$.

The next step is to construct such a vector that contains the
physical fields under the equations \p{l0l1}. The constraint on this
vector, stipulated by the operator $L_1^+$, is not imposed. The BRST
procedure for such systems was studied in papers \cite{BuchKrP} --
\cite{Buchbinder:2015kca} and we apply it to the infinite spin
system under consideration.

Thus, using the operators $F_a = (T_0,L_0,L_1,L_1^+$) and the
corresponding
 ghosts $c_a = (q_0, \eta_0, \eta_1, \eta_1^+)$ we construct Hermitian
BRST charge $Q=Q^\dagger$ in the form
\begin{equation}\label{Q}
Q\ =\ q_0T_0+\eta_0L_0+\eta_1^+L_1+\eta_1L_1^+
+\eta_1^+\eta_1(N+\bar{N}+2)\mathcal{P}_0-q_0^2\,\mathcal{P}_0\,,
\end{equation}
which is nilpotent by definition
\begin{equation}\label{Q2}
Q^2 =0\,.
\end{equation}
The BRST-charge acts in the extended Fock space, where the action of
fermionic $\eta_0$, $\eta_1$, $\eta_1^+$ and bosonic $q_0$ ghost
``coordinates'', as well as the corresponding
 ghost ``momenta'' $\mathcal{P}_0$, $\mathcal{P}_1^+$, $\mathcal{P}_1$
  and $q_0$, are defined earlier.
 These ghost operators obey the (anti)commutation relations
\begin{equation}\label{ghosts}
[q_0,p_0]=i\,,
\qquad
\{\eta_1,{\cal{}P}_1^+\}
= \{{\cal{}P}_1, \eta_1^+\}
=\{\eta_0,{\cal{}P}_0\}
=1
\end{equation}
and act on the "vacuum" vector as follows
\begin{equation}
p_0|0\rangle=
\eta_1|0\rangle=\mathcal{P}_1|0\rangle=\mathcal{P}_0|0\rangle=0\,.
\end{equation}
They possess the standard  ghost numbers,
$gh(\mbox{``coordinates''}) =  - gh(\mbox{``momenta''}) = 1$,
providing the property  $gh(Q) = 1$.

The operator (\ref{Q}) acts in the extended Fock space of the vectors
\begin{equation}\label{ext-vector}
|\Phi_C\rangle \ = \ |\varphi_C\rangle+\eta_0\mathcal{P}_1^+|\varphi_{1C}\rangle
+\eta_1^+\mathcal{P}_1^+|\varphi_{2C}\rangle
+q_0\mathcal{P}_1^+|\varphi_{3C}\rangle \,.
\end{equation}
The equation of motion of this BRST-field is postulated in the form
\begin{equation}\label{Q-eq}
Q_C{}^D \; |\Phi_D\rangle=0 \,.
\end{equation}
Due to the nilpotency of the BRST charge the field \p{ext-vector} is defined up to the gauge transformations
\begin{equation}\label{Q-var}
|\Phi'_C\rangle = |\Phi_C\rangle +Q_C{}^D \; |\Lambda_D\rangle \,,
\end{equation}
where the gauge parameter $|\Lambda_D\rangle$ has (since $gh(Q) = 1$
and $gh(\mathcal{P}_1^+) = - 1$) the form
\begin{equation}
|\Lambda_D\rangle=\mathcal{P}_1^+|\lambda_D\rangle \,.
\label{g-parameter}
\end{equation}
The fields $|\varphi_C\rangle$, $|\varphi_{1C}\rangle$,
$|\varphi_{2C}\rangle$, $|\varphi_{3C}\rangle$ and the gauge
parameter  $|\lambda_C\rangle$ in (\ref{ext-vector}) and
(\ref{g-parameter}) have the decompositions similar with
$|\varphi_C\rangle$ in (\ref{GFState}).

We emphasize that we take ``the momentum representation''
 with respect to the canonical pair of ghost variables
$(\eta_1,\mathcal{P}_1^+)$, in contrast to ``the coordinate
representation'' for other canonical pairs of ghosts. This
prescription leads to the possibility to consider the corresponding
constraint $L_1^+$ by using gauge symmetry, as it is given in
Appendix\,B. Description of such a treatment to use the constraints
in the BRST approach was given in \cite{AlkGr,ACG18}.

The equation of motion $Q|\Phi\rangle=0$ \p{Q-eq} can be rewritten
in term of the vectors $|\varphi\rangle$, $|\varphi_i\rangle$,
$i=1,2,3$ in the form (we omit here the Dirac indices $C,...$ in all
quantities)
\begin{eqnarray}
\label{eq1}&&
T_0|\varphi\rangle +(l_1^+-\mu)|\varphi_{3}\rangle=0\,,
\\ [5pt]
\label{eq2}&&
l_0|\varphi\rangle-(l_1^+-\mu)|\varphi_{1}\rangle=0\,,
\\ [5pt]
\label{eq3}&&
(l_1-\mu)|\varphi\rangle+(N+\bar{N}+2)|\varphi_{1}\rangle-
(l_1^+-\mu)|\varphi_{2}\rangle=0\,,
\\ [5pt]
\label{eq4}&&
T_0|\varphi_{1}\rangle+l_0|\varphi_{3}\rangle=0\,,
\\ [5pt]
\label{eq5}&&
T_0|\varphi_{2}\rangle+(l_1-\mu)|\varphi_{3}\rangle=0\,,
\\ [5pt]
\label{eq6}&&
-(l_1-\mu)|\varphi_{1}\rangle+l_0|\varphi_{2}\rangle=0\,,
\\ [5pt]
\label{eq7}&&
-|\varphi_{1}\rangle-T_0|\varphi_{3}\rangle=0\,.
\end{eqnarray}
In this case the gauge transformations
$\delta|\Phi\rangle=Q|\Lambda\rangle$ \p{Q-var} look like
\begin{equation}\label{var-BRST}
\delta|\varphi\rangle=(l_1^+-\mu)|\lambda\rangle \,,\qquad
\delta|\varphi_{1}\rangle=l_0|\lambda\rangle \,,\qquad
\delta|\varphi_{2}\rangle=(l_1-\mu)|\lambda\rangle \,,\qquad
\delta|\varphi_{3}\rangle=-T_0|\lambda\rangle \,.
\end{equation}

Making use of equation \p{eq7}, one can express field
$|\varphi_{1}\rangle$ in the form
$|\varphi_{1}\rangle=-T_0|\varphi_{3}\rangle$ and then substitute it
to other equations. As a result we obtain only three independent
equations
\begin{eqnarray}
\label{Eq1}
&&
T_0|\varphi\rangle+(l_1^+-\mu)|\varphi_{3}\rangle=0\,,
\\ [5pt]
\label{Eq2}
&&
(l_1-\mu)|\varphi\rangle-(N+\bar{N}+2)T_0|\varphi_{3}\rangle-
(l_1^+-\mu)|\varphi_{2}\rangle=0\,,
\\ [5pt]
\label{Eq3}
&&
T_0|\varphi_{2}\rangle+(l_1-\mu)|\varphi_{3}\rangle=0\,.
\end{eqnarray}
Residual gauge transformations, which follow from \p{var-BRST}, have
the form
\begin{equation}\label{var-BRSTa}
\delta|\varphi\rangle=(l_1^+-\mu)|\lambda\rangle \,,\qquad
\delta|\varphi_{2}\rangle=(l_1-\mu)|\lambda\rangle \,,\qquad
\delta|\varphi_{3}\rangle=-T_0|\lambda\rangle \,.
\end{equation}

\subsection{Construction of the Lagrangian}

It is easy to see that the equations \p{Eq1}--\p{Eq3} are Lagrangian
equations for the following Lagrangian
\begin{equation}
\begin{array}{rcl}
\mathcal{L}
&=&
\langle\bar{\varphi}|\Bigl\{T_0|\varphi\rangle+
(l_1^+-\mu)|\varphi_{3}\rangle\Bigr\}
-\langle\bar{\varphi}_2|\Bigl\{T_0|\varphi_{2}\rangle+
(l_1-\mu)|\varphi_{3}\rangle\Bigr\}
\\ [6pt]
&+&{}
 \langle\bar{\varphi}_3|\Bigl\{(l_1-\mu)|\varphi\rangle-
(N+\bar{N}+2)T_0|\varphi_{3}\rangle
-(l_1^+-\mu)|\varphi_{2}\rangle\Bigr\}  \; .
\label{LagrFock}
 \end{array}
\end{equation}
Then we calculate the inner products in \eqref{LagrFock}. After that
we convert the Weyl spinor indices of the component fields into
vector ones, leaving Dirac indices intact. This way we obtain the
following tensor Dirac spinor fields
\begin{equation}\label{phi-vec-n}
\varphi_{A\,m(s)} :=
\varphi_{A\,m_1\ldots m_s} \,,
\end{equation}
which are symmetric with respect to all vector indices and defined by
\begin{equation}\label{phi-vec-sp}
\varphi_{A\,m(s)} =
\frac{(-1)^s}{2^s}\;
\tilde{\sigma}_{m_1}^{\dot\beta_1\alpha_1}\ldots\tilde{\sigma}_{m_s}^{\dot\beta_s\alpha_s}
\;\varphi_{A\,\alpha(s)\dot\beta(s)} \,.
\end{equation}
We use this Rarita-Schwinger-like fields in the expansions of all
``physical'' fields $\varphi$ and gauge field $\lambda$. By
construction all ``physical'' and gauge fields are totally symmetric
traceless tensor Dirac spinors
\begin{equation}\label{TraceLess}
\eta^{m_1m_2}\varphi_{A\,m(s)}=0\,.
\end{equation}
One can check that
\begin{eqnarray}
&&
\langle\bar{\varphi}^A_{s}|\chi_{A\,s}\rangle
=(-1)^s\;\bar{\varphi}^{A\,\alpha(s)\dot\beta(s)}\chi_{A\,\alpha(s)\dot\beta(s)}
=2^s\bar{\varphi}^{A\,m(s)}\chi_{A\,m(s)}\,,
\\ [5pt]
&&
\langle\bar{\varphi}^A_{s}|l_1|\chi_{A\,s+1}\rangle
=2^s\,\bar{\varphi}^{A\,m(s)}(-2i)(s+1)\partial^n\chi_{A\,nm(s)}\,,
\\ [5pt]
&&
\langle\bar{\varphi}^A_{s}|l_1^+|\chi_{A\,s-1}\rangle
=2^s\,\bar{\varphi}^{A\,m(s)}(-is)\partial_{m_s}\chi_{A\,m(s-1)}\,.
\end{eqnarray}
As a result the BRST Lagrangian \p{LagrFock} yields the following
component Lagrangian
\begin{eqnarray}
\mathcal{L}
&=&
\sum_{s=0}^\infty 2^s \;\bar{\varphi}^{m(s)}\Bigl[
i\! \not\!\partial\,\varphi_{m(s)}-is\,\partial_{m_s} \varphi_{3\,m(s-1)}
-\mu\,\varphi_{3\,m(s)}
\Bigr]
\nonumber
\\
&&{}
-\sum_{s=0}^\infty 2^s \;\bar{\varphi}_{2}^{m(s)}\Bigl[
i\! \not\!\partial\,\varphi_{2m(s)}
-2i(s+1)\partial^n\varphi_{3\,nm(s)}
-\mu\, \varphi_{3\,m(s)}
\Bigr]
\label{Lagr-comp}
\\
&&
+\sum_{s=0}^\infty 2^s \;\bar{\varphi}^{m(s)}_3\Bigl[
-2i(s+1)\partial^n\varphi_{nm(s)}
-\mu \varphi_{m(s)}
\nonumber
\\
&&\hspace{15ex}{}
+is\,\partial_{m_s} \varphi_{2\,m(s-1)}
+\mu \varphi_{2\,m(s)}
-2i(s+1)\!\!\not\!\partial\,\varphi_{3\,m(s)}
\Bigr]\,.
\nonumber
\end{eqnarray}
Gauge transformations \p{var-BRSTa} take the form
\begin{eqnarray}\label{var1+}
\delta\varphi_{m(s)}&=&-is\,\partial_{(m_s}\lambda_{m(s-1))}
+\frac{i(s-1)}{2}\,\eta_{(m_{s-1}m_{s}}\,\partial^n\lambda_{m(s-2))n}
-\mu\,\lambda_{m(s)}
\\ [5pt]
\delta\varphi_{3\,m(s)}&=&-i\!\!\not\!\partial\,\lambda_{m(s)}
\label{dphi3}
\label{var2+}\\ [5pt]
\delta\varphi_{2\,m(s)}&=&-2i(s+1)\,\partial^{n}\lambda_{nm(s)}-\mu\,\lambda_{m(s)} \,.\label{var3+}
\end{eqnarray}
The relations (\ref{Lagr-comp}) and (\ref{var1+})--(\ref{var3+}) are the
final results.

Let us show that the Lagrangian (\ref{Lagr-comp}) reproduces the
conditions (\ref{C1+}) and (\ref{C2+}) after the appropriate gauge
fixing. Equations of motion following from Lagrangian
(\ref{Lagr-comp})
in component form have the form
\begin{eqnarray}
&&
i\!\!\not\!\partial\,\varphi_{m(s)}
-is\,\partial_{(m_s} \varphi_{3\,m(s-1))}
+\frac{i(s-1)}{2}\,\eta_{(m_{s-1}m_s}\partial^n\varphi_{3\,m(s-2))n}
-\mu\,\varphi_{3\,m(s)}
=0\,,
\label{Eq1+}
\\ [5pt]
&&{}
i\!\!\not\!\partial\,\varphi_{2m(s)}
-2i(s+1)\partial^n\varphi_{3\,nm(s)}
-\mu\, \varphi_{3\,m(s)}
=0\,,
\label{Eq2+}
\\ [5pt]
&&
-2i(s+1)\partial^n\varphi_{nm(s)}
-\mu \varphi_{m(s)}
+is\,\partial_{(m_s} \varphi_{2\,m(s-1))}
-\frac{i(s-1)}{2}\,\eta_{(m_{s-1}m_s}\partial^n\varphi_{2\,m(s-2))n}
\label{Eq3+}
\\ [5pt]
&&\hspace{8cm}{}
+\mu \varphi_{2\,m(s)}
-2i(s+1)\!\!\not\!\partial\,\varphi_{3\,m(s)}
=0\,.
\nonumber
\end{eqnarray}
Here we taken into account that the equations of motion are
constructed for the traceless fields. We can remove the fields
$\varphi_{3\,m(s)}$ using their gauge transformations and after that
we can make the gauge transformations using restricted gauge
parameters subjected to the conditions
$\not\!\partial\,\lambda_{m(s)}=0$. Note that the equations
\eqref{Eq2+} on fields $\varphi_{2\,m(s)}$ take the same form as the
equations on the gauge parameters
$\not\!\partial\,\varphi_{2\,m(s)}=0$. Therefore we have enough
gauge freedom to remove fields $\varphi_{2\,m(s)}$. Thus, after
removing the fields $\varphi_{2\,m(s)}$ and $\varphi_{3\,m(s)}$, the
equations of motion \eqref{Eq1+}, \eqref{Eq3+}  take the form
\begin{equation}
\not\!\partial\,\varphi_{m(s)}=0\,, \qquad
-2i(s+1)\partial^n\varphi_{nm(s)}-\mu \varphi_{m(s)} =0\,
\end{equation}
and coincide with (\ref{C1+}) and (\ref{C2+}). As a result, we have
shown that the Lagrangian  \eqref{Lagr-comp} describes the fermionic
infinite spin field. We emphasize that this Lagrangian
(\ref{Lagr-comp}) has consistently derived in the framework of the
general BRST construction. In fact, it is a direct consequence of
the basic conditions (\ref{psi-eq0})-(\ref{psi-eq3}). The only
assumption we made was a homogeneity condition (\ref{psi-eq3}) (see
also \p{phi-eq3-a}) for the fields $\Phi_\alpha(x,\xi,\bar\xi)$ and
$\bar{\mathcal{X}}^{\,\dot\alpha}(x,\xi,\bar\xi)$ in
(\ref{ISF-exp}).

We see that the Lagrangian (\ref{Lagr-comp}) depends on three sets
of traceless Dirac fields $\varphi_{m(s)}$, $\varphi_{2\,m(s)}$,
$\varphi_{3\,m(s)}$ \eqref{TraceLess} and each traceless field can
be decomposed into two $\gamma$-traceless fields thus Lagrangian
\eqref{Lagr-comp} depends on six sets of Dirac $\gamma$-traceless
fields. We emphasize that just such a Lagrangian corresponds to the
fields satisfying the basic conditions
(\ref{psi-eq0})-(\ref{psi-eq3}).

Recently the Lagrangian for fermionic infinite spin field has been
proposed in  \cite{Mets17}, \cite{HabZin} by combining the free
massless fermionic fields with definite helicities and assuming the
special gauge symmetry. This Lagrangian depends on one set of Dirac
triple $\gamma$-traceless fields and each field also can be
decomposed into three Dirac $\gamma$-traceless fields. Thus, one can
say that the set of the fields of our Lagrangian (\ref{Lagr-comp})
(as well as gauge parameters and, respectively, degrees of freedom)
is twice as large as that of Lagrangian proposed in \cite{Mets17},
\cite{HabZin}.
Nevertheless we emphasize once more that Lagrangian
(\ref{Lagr-comp}) was consistently derived only on the base of the
basic conditions (\ref{psi-eq0})-(\ref{psi-eq3}) including the
homogeneity condition.
At present it is not clear how our Lagrangian (\ref{Lagr-comp})
relates to the Lagrangian obtained in the works \cite{Mets17},
\cite{HabZin}.

\section{Summary and outlook}
We have constructed the Lagrangian for the infinite half-integer spin fields.
This construction is characterized by the following:
\begin{itemize}
\item
Irreducible infinite half-integer spin representation is described
by the fields \p{ISF-exp}, which depend on additional even spinor
variables. The fields \p{ISF-exp} contain the fields \p{Phi-expan},
\p{bPhi-expan} that satisfy the conditions \p{cond-1}, \p{cond-2}
and have the power expansion.

\item
The second Casimir operator (\ref{W2-ex}), which acts in the space
of
 infinite spin fields with additional
 spinor variables is derived.

\item
Without the presence of the  $\delta$-function in \p{ISF-exp}, the
fields \p{Phi-expan}, \p{bPhi-expan} with the component field
equations \p{C1+}, \p{C2+} describe reducible infinite half-integer
spin representation.

\item
The fermionic infinite spin equations (\ref{C1+}), (\ref{C2+}) are
reproduced by the constraints \p{l0l1} imposed on the vector
\p{GFState}.
\item
The constraints \p{l0l1} are obtained from the BRST equation
\p{Q-eq} for the vector \p{ext-vector}, where BRST operator is
defined in \p{Q}. After the elimination of some auxiliary states, we
stay with the physical and gauge states, that are described by the
equations of motion \p{Eq1}, \p{Eq2}, \p{Eq3} and the gauge
transformations \p{var-BRSTa}.
\item
The Lagrangian \p{Lagr-comp} is invariant under the gauge
transformations \p{var1+}, \p{var2+}, \p{var3+} of component fields.
The corresponding equations of motion of the component fields have
the form \p{Eq1+}, \p{Eq2+}, \p{Eq3+}.

\end{itemize}

Let us note some comments on the constructed Lagrangian.
\begin{description}
\item[i)]
It is interesting to generalize the BRST approach for obtaining the
Lagrangian for the irreducible representation of infinite spin
field.
\item[ii)]
It is interesting to generalize the BRST approach for obtaining
field Lagrangian to supersymmetric infinite spin field theory.
\item[iii)]
It would also be interesting to obtain the Lagrangian of such type
for the infinite spin fields in the AdS space.
\end{description}

\section*{Acknowledgments}

Authors would like to thank Konstantin Alkalaev, Ruslan Metsaev and
Yurii Zinoviev for useful discussions and valuable comments. I.L.B.
and V.A.K. acknowledge the support of the Russian Foundation for
Basic Research, project No.\,18-02-00153. A.P.I. acknowledges the
support of the Russian Science Foundation, grant No.\,19-11-00131.

\section*{Appendix\,A: \ Notations }
\def\theequation{A.\arabic{equation}}
\setcounter{equation}0

\quad\,
In this Appendix we present the notations used in this paper.

The space-time metric is $\eta_{mn}={\rm diag}(-1,+1,+1,+1)$. The
totally antisymmetric tensor $\varepsilon_{mnkl}$ has the component
$\varepsilon_{0123}=-1$. The two-component Weyl spinor indices are
raised and lowered by $\epsilon_{\alpha\beta}$,
$\epsilon^{\alpha\beta}$, $\epsilon_{\dot\alpha\dot\beta}$,
$\epsilon^{\dot\alpha\dot\beta}$ with the non-vanishing components
$\epsilon_{12}=-\epsilon_{21}=\epsilon^{21}=-\epsilon^{12}=1$:
\begin{equation}
\label{up-sp}
\psi_\alpha=\epsilon_{\alpha\beta}\psi^\beta\,,\qquad \psi^\alpha=\epsilon^{\alpha\beta}\psi_\beta \, ,
\end{equation}
etc. Relativistic
$\sigma$-matrices are
\begin{equation}
\label{sigma-m}
(\sigma_m)_{\alpha\dot\beta}=({\bf 1_2};\sigma_1,\sigma_2,\sigma_3)_{\alpha\dot\beta} \, ,
\end{equation}
where $\sigma_1,\sigma_2,\sigma_3$ are the Pauli matrices. The matrices
\begin{equation}
\label{t-sigma-m}
(\tilde\sigma_{m})^{\dot\alpha\beta}=\epsilon^{\dot\alpha\dot\delta}\epsilon^{\beta\gamma}
(\sigma_m)_{\gamma\dot\delta}=({\bf 1_2};-\sigma_1,-\sigma_2,-\sigma_3)^{\dot\alpha\beta}
\end{equation}
satisfy the relations
\begin{equation}
\label{sigma-eq}
\sigma^m_{\alpha\dot\gamma}\tilde\sigma^{n\,\dot\gamma\beta}+\sigma^m_{\alpha\dot\gamma}\tilde\sigma^{n\,\dot\gamma\beta}
=-2\,\eta^{mn}\delta^\beta_\alpha\,,\qquad \sigma^m_{\alpha\dot\beta}\tilde\sigma_n^{\dot\beta\alpha}=-2\,\delta^m_n\, .
\end{equation}
The link between the Minkowski four-vector $A_m$ and bi-spinor $A_{\alpha\dot\beta}$ is
given by
$A_{\alpha\dot\beta}= A_m(\sigma^m)_{\alpha\dot\beta}$,
$A^{\dot\alpha\beta}= A_m(\tilde\sigma^m)^{\dot\alpha\beta}$,
$A_m
=-\frac{1}{2}\,A_{\alpha\dot\beta}(\tilde\sigma_m)^{\dot\beta\alpha}$,
so that $ A^m B_m =-\frac{1}{2}\,A_{\alpha\dot\beta}B^{\dot\beta\alpha}$.

The $\sigma$-matrices with two vector indices are defined by
\begin{equation}
\label{sigma-mn}
(\sigma_{mn})_{\alpha}{}^{\beta}=-\frac14\,(\sigma_m \tilde\sigma_{n}-\sigma_n \tilde\sigma_{m})_{\alpha}{}^{\beta}\,,\qquad
(\tilde\sigma_{mn})^{\dot\alpha}{}_{\dot\beta}=-\frac14\,(\tilde\sigma_m \sigma_{n}-\tilde\sigma_n \sigma_{m})^{\dot\alpha}{}_{\dot\beta}\,.
\end{equation}
They satisfy the identities
\begin{equation}
\label{eps-sigma-mn}
\varepsilon^{mnkl}\sigma_{kl}=-2i\,\sigma^{mn}\,,\qquad
\varepsilon^{mnkl}\tilde\sigma_{kl}=2i\,\tilde\sigma^{mn}\,.
\end{equation}
Using the $\sigma$-matrices \p{sigma-mn}, we represent the
antisymmetric second rank vector tensor in the form
\begin{equation}
\label{X-sigma-mn}
X_{mn}=X_{[mn]}=(\sigma_{mn})_{\alpha\beta}X^{\alpha\beta} -
(\tilde\sigma_{mn})_{\dot\alpha\dot\beta}X^{\dot\alpha\dot\beta} \,,
\end{equation}
where the inverse expressions for the symmetric second rank spinor
tensors are
\begin{equation}
\label{X-sigma-mn-2}
X_{\alpha\beta}=X_{(\alpha\beta)}=\frac12\,(\sigma^{mn})_{\alpha\beta}X_{mn}  \,,\qquad
X_{\dot\alpha\dot\beta}=X_{(\dot\alpha\dot\beta)}= -\frac12\, (\tilde\sigma^{mn})_{\dot\alpha\dot\beta}X_{mn} \,.
\end{equation}

In the Weyl representation, the Dirac matrices $(\gamma_m)_A{}^B$,
$A,B=1,2,3,4$ have the form
\begin{equation}
\label{gamma-matr}
\gamma_m=
\left(
  \begin{array}{cc}
    0 & (\sigma_m)_{\alpha\dot\beta} \\
    (\tilde\sigma_m)^{\dot\alpha\beta} & 0 \\
  \end{array}
\right)
\,, \qquad \{\gamma_m,\gamma_n\}=-2\eta_{mn}\,.
\end{equation}
We use the following notations:
\begin{equation}
\label{nabla-gamma}
\not\!{P}_A{}^B:=P_m(\gamma_m)_A{}^B\,,\qquad  \not\!\partial_A{}^B:=\partial_m(\gamma_m)_A{}^B\,.
\end{equation}

Four-component Dirac spinor $\Psi_A$ is represented by two Weyl
spinors
\begin{equation}
\label{Dir-sp}
\Psi_A=\left(
\begin{array}{c}
\psi_\alpha \\ [5pt]
\bar\chi^{\dot\alpha} \\
\end{array}
\right)
\,.
\end{equation}
Dirac conjugate spinor $\bar\Psi=\Psi^\dagger\gamma_0$ has the
components
\begin{equation}
\label{cDir-sp}
\bar\Psi^A=\left(\chi^\alpha ,\bar\psi_{\dot\alpha} \right),\qquad \bar\psi_{\dot\alpha}=(\psi_\alpha)^*\,,\quad
\chi^{\alpha}=(\bar\chi^{\dot\alpha})^*
\,.
\end{equation}
In case of the Majorana spinor, the equality
$\chi_{\alpha}=\psi_\alpha$ holds.

\section*{Appendix\,B: \ Free infinite spin fields with additional spinor coordinates in the space-time description}
\def\theequation{B.\arabic{equation}}
\setcounter{equation}0

To analyze the field contents let us consider the light-cone
reference system, where $p^+:=p^0+p^3=2E$, $p^-:=p^0-p^3=0$,
$p^1=p^2=0$ and the four-momentum  has the form
\begin{equation}\label{sys-st-m}
p^m=(E,0,0,E)\,,
\qquad
p_{\alpha\dot{\beta}}=\begin{pmatrix} p_{1\dot{1}}& 0 \\ 0 & 0 \end{pmatrix}=\begin{pmatrix} 2E& 0 \\ 0 & 0 \end{pmatrix}\,,
\quad
p^{\dot{\alpha}\beta}=\begin{pmatrix} 0& 0 \\ 0 & p^{\dot{2}2} \end{pmatrix}=\begin{pmatrix} 0& 0 \\ 0 & 2E \end{pmatrix}
\,.
\end{equation}
In this system the helicity operator takes the form
\begin{equation}
h=\frac{W^0}{E}=-\frac{1}{2}\,\varepsilon_{0mnk}J^{mn}\frac{P^k}{E}
=\frac{i}{2}\,\varepsilon_{0mn3}M^{mn}
=(\sigma_{03})^{\alpha\beta}M_{\alpha\beta}
+(\tilde{\sigma}_{03})^{\dot{\alpha}\dot{\beta}}\bar{M}_{\dot{\alpha}\dot{\beta}}\,.
\end{equation}
Taking into account the relations (see \p{M}, \p{M-sp})
\begin{equation}
\begin{array}{rcl}
(\sigma_{30})_{\alpha\beta}M^{\alpha\beta}
&=&
{\displaystyle -\frac{1}{2}\Bigl(\xi^1\frac{\partial}{\partial\xi^1}-\xi^2\frac{\partial}{\partial\xi^2}
+\zeta^1\frac{\partial}{\partial\zeta^1}-\zeta^2\frac{\partial}{\partial\zeta^2}\Bigr)\,,}
\\ [7pt]
(\tilde{\sigma}_{30})_{\dot{\alpha}\dot{\beta}}\bar{M}^{\dot{\alpha}\dot{\beta}}
&=&
{\displaystyle \frac{1}{2}\Bigl(\bar\xi^{\dot{1}}\frac{\partial}{\partial\bar\xi^{\dot{1}}}-\bar\xi^{\dot{2}}\frac{\partial}{\partial\bar\xi^{\dot{2}}}
+\bar\zeta^{\dot{1}}\frac{\partial}{\partial\bar\zeta^{\dot{1}}}-\bar\zeta^{\dot{2}}\frac{\partial}{\partial\bar\zeta^{\dot{2}}}\Bigr)\,,}
\end{array}
\end{equation}
we obtain the following expression for helicity operator in the light-cone system
\begin{equation}\label{h}
h=\frac{1}{2}\Bigl(\xi^1\frac{\partial}{\partial\xi^1}-\xi^2\frac{\partial}{\partial\xi^2}
-\bar\xi^{\dot{1}}\frac{\partial}{\partial\bar\xi^{\dot{1}}}+\bar\xi^{\dot{2}}\frac{\partial}{\partial\bar\xi^{\dot{2}}}
+\zeta^1\frac{\partial}{\partial\zeta^1}-\zeta^2\frac{\partial}{\partial\zeta^2}
-\bar\zeta^{\dot{1}}\frac{\partial}{\partial\bar\zeta^{\dot{1}}}+\bar\zeta^{\dot{2}}\frac{\partial}{\partial\bar\zeta^{\dot{2}}}\Bigr)
\,.
\end{equation}
Note, to obtain this expression for helicity operator we do not use
the equations of motion.

\subsection*{B.1 \ Integer spins}
\subsubsection*{B.1.1 \ Field without $\delta$-function}
Let us consider the generalized field in momentum representation
\begin{equation} \label{Phi-expan0}
\Phi(p,\xi,\bar{\xi}) =
\sum_{s=0}^\infty
\frac{1}{s!}\;\varphi_{\alpha_1\ldots \alpha_s\dot\beta_1\ldots\dot\beta_s}(p)\;
\xi^{\alpha_1}\ldots\xi^{\alpha_s}\;\bar{\xi}^{\dot\beta_1}\ldots\bar{\xi}^{\dot\beta_s} \,,
\end{equation}
which satisfy the equations of motion
\begin{equation}\label{cond-10}
P^{2}\Phi =0\,,\qquad
\left(\frac{\partial}{\partial \bar\xi^{\dot\alpha}} P^{\dot\alpha\beta}\frac{\partial}{\partial \xi^{\beta}}+\mu\right)\Phi =0\,.
\end{equation}
For the component fields
\begin{equation} \label{short-notat0}
\varphi_{ \alpha(k)\dot\beta(s)}:=\varphi_{ \alpha_1\ldots
\alpha_s\dot\beta_1\ldots\dot\beta_s}\,
\end{equation}
these equations have the form
\begin{equation}
p^{2}\varphi_{\alpha(s)\dot\beta(s)}=0\,,\qquad
s\,p^{\dot\beta_s\alpha_s}\varphi_{\alpha(s)\dot\beta(s)}=-\mu\varphi_{\alpha(s-1)\dot\beta(s-1)}.
\label{C1+0}
\end{equation}

First equation in (\ref{C1+0}) is massless Klein-Gordon equation,
which can be written in the light-cone system (\ref{sys-st-m}).
Second set of the equations in (\ref{C1+0})
$$
p^{\dot\beta\alpha}\varphi_{\alpha\dot\beta}=- \mu \varphi_0\,,\quad
p^{\dot\beta\alpha}\varphi_{\alpha\alpha_1\dot\beta\dot\beta_1}=- \frac{\mu}{2}\, \varphi_{\alpha_1\dot\beta_1}\,,\quad
p^{\dot\beta\alpha}\varphi_{\alpha\alpha_1\alpha_2\dot\beta\dot\beta_1\dot\beta_2}=- \frac{\mu}{3}\, \varphi_{\alpha_1\alpha_2\dot\beta_1\dot\beta_2}\,,\quad \ldots
$$
in the system (\ref{sys-st-m}) take the form
$$
2E\,\varphi_{2\dot2}=- \mu \varphi_0\,,\quad
2E\,\varphi_{(2\alpha_1)(\dot2\dot\beta_1)}=- \frac{\mu}{2}\, \varphi_{\alpha_1\dot\beta_1}\,,\quad
2E\,\varphi_{(2\alpha_1\alpha_2)(\dot2\dot\beta_1\dot\beta_2)}=- \frac{\mu}{3}\, \varphi_{\alpha_1\alpha_2\dot\beta_1\dot\beta_2}\,,\quad \ldots
$$
As we see, the independent fields (for below fields we point out
their helicities, calculated by formula (\ref{h})) are
\begin{equation}\label{exp0}
\underbrace{\varphi_0}_{0}\,,\qquad \underbrace{\varphi_{1\dot1}}_{0}\,,\underbrace{\varphi_{1\dot2}}_{1}\,,\underbrace{\varphi_{2\dot1}}_{-1}\,,\qquad
\underbrace{\varphi_{11\dot1\dot1}}_{0}\,,\underbrace{\varphi_{11\dot1\dot2}}_{1}\,,\underbrace{\varphi_{12\dot1\dot1}}_{-1}\,,
\underbrace{\varphi_{11\dot2\dot2}}_{2}\,,\underbrace{\varphi_{22\dot1\dot1}}_{-2}\,,\qquad \ldots
\end{equation}
We indicate the structure of this set of states: on $s$-th step in
the expansion, the physical states have the helicities from $0$ to
$\pm s$, on $(s+1)$-th step they have the helicities from $0$ to
$\pm (s+1)$ etc. That is, at each step there arise the states with
the same helicities  as in the previous step, and the additional
states with helicities which are one more modulo larger. So, in the
spectrum, there are all helicities, and there is an infinite number
of states with an arbitrary fixed helicity. That is, this
representation of the infinite spin is not irreducible, it is
infinitely degenerate.

We get the same result using a slightly different procedure.

In the light-cone system
(\ref{sys-st-m}) second equation in \p{cond-10}
\begin{equation}\label{cond-10-2}
\left(\frac{\partial}{\partial \xi^{2}}\frac{\partial}{\partial \bar\xi^{\dot2}} +\frac{\mu}{2E}\right)
\Phi(E;\xi^{1},\xi^{2},\bar\xi^{\dot1},\bar\xi^{\dot2}) =0
\end{equation}
has the general polynomial solution
\begin{eqnarray}\label{cond-10-2a}
\Phi(\xi^{1},\xi^{2},\bar\xi^{\dot1},\bar\xi^{\dot2}) &=&
G^{(0)}(E;\xi^{2},\bar\xi^{\dot2})\
F^{(0)}(E;\xi^{1},\bar\xi^{\dot1}) \\ [7pt]
&&
+
\sum_{n=1}^{\infty}(\xi^{2})^n\, G^{(n)}(E;\xi^{2},\bar\xi^{\dot2})\ F^{(-n)}(E;\xi^{1},\bar\xi^{\dot1})
\nonumber \\ [7pt]
&&
+
\sum_{n=1}^{\infty}(\bar\xi^{\dot2})^n\, G^{(n)}(E;\xi^{2},\bar\xi^{\dot2})\ F^{(n)}(E;\xi^{1},\bar\xi^{\dot1})\,.
\nonumber
\end{eqnarray}
where the multipliers  are modified Bessel functions \cite{BeitmanErd}
\begin{equation}\label{cond-10-2b}
G^{(0)}(E;\xi^{2},\bar\xi^{\dot2})\ :=\ I_0\left(\sqrt{-\frac{2\mu\,\xi^{2}\bar\xi^{\dot2}}{E}}\right)\ = \
\sum_{k=0}^{\infty}\frac{1}{(k!)^2}\left( -\frac{\mu\,\xi^{2}\bar\xi^{\dot2}}{2E}\right)^k
\end{equation}
and the polynomial functions with respect to the variable $\xi^{2}$:
\begin{equation}\label{cond-10-2c}
G^{(n)}(E;\xi^{2},\bar\xi^{\dot2})\
:=\
\sum_{k=0}^{\infty}\frac{1}{k!(k+n)!}\left( -\frac{\mu\,\xi^{2}\bar\xi^{\dot2}}{2E}\right)^k\,.
\end{equation}
In the case of a general field \p{Phi-expan0} with zero degree of
homogeneity, the fields $F^{(k)}(E;\xi^{1},\bar\xi^{\dot1})$,
$-\infty<k<\infty$ in the expansion \p{cond-10-2a} have the
following degrees of homogeneity ($U(1)$-charges):
\begin{equation}\label{cond-10-2d}
\left(\xi^{1}\frac{\partial}{\partial\xi^{1}}-\bar\xi^{\dot1}\frac{\partial}{\partial\bar\xi^{\dot1}} \right)F^{(k)}(E;\xi^{1},\bar\xi^{\dot1})=
k\,F^{(k)}(E;\xi^{1},\bar\xi^{\dot1}) \,.
\end{equation}
As result, the solutions of the equations \p{cond-10-2d} have the
form
\begin{equation}\label{cond-10-2e}
F^{(k)}(E;\xi^{1},\bar\xi^{\dot1})=\left\{
\begin{array}{l}
(\xi^{1})^k {\displaystyle \sum\limits_{l=0}^{\infty}} (\xi^{1}\bar\xi^{\dot1})^l f^{(k)}_l(E)\,,\quad k\geq0\,,\\ [6pt]
(\bar\xi^{\dot1})^{-k}{\displaystyle \sum\limits_{l=0}^{\infty}}(\xi^{1}\bar\xi^{\dot1})^l f^{(k)}_l(E)\,,\quad k<0 \,.
\end{array}
\right.
\end{equation}
In these expansions, the infinite number of the functions
$f^{(k)}_l(E)$, $l=0,1,\ldots,\infty$ describe the infinite number
of massless states with helicities $k$. In particle, the
helicity-zero fields $f^{(0)}_l$, $l=0,1,\ldots,\infty$ correspond
to the fields $\varphi_{0}$, $\varphi_{1\dot1}$,
$\varphi_{11\dot1\dot1}, \ldots$ in \p{exp0},  helicity-one fields
$f^{(1)}_l$, $l=0,1,\ldots,\infty$ correspond to the fields
$\varphi_{1\dot2}$, $\varphi_{11\dot1\dot2}$,
$\varphi_{111\dot1\dot1\dot2}, \ldots$ in \p{exp0},
helicity-minus-one fields $f^{(-1)}_l$, $l=0,1,\ldots,\infty$
correspond to the fields $\varphi_{2\dot1}$,
$\varphi_{12\dot1\dot1}$, $\varphi_{112\dot1\dot1\dot1}, \ldots$ in
\p{exp0}, etc.

\subsubsection*{B.1.2 \ Field with $\delta$-function}

Now we consider the field
\begin{equation} \label{Phi-expan0d}
\Psi(p,\xi,\bar{\xi}) =
\delta(\xi p\bar\xi-\mu)\Phi(p,\xi,\bar{\xi}) =
\delta(\xi p\bar\xi-\mu)\sum_{s=0}^\infty
\frac{1}{s!}\;\varphi_{\alpha_1\ldots \alpha_s\dot\beta_1\ldots\dot\beta_s}(p)\;
\xi^{\alpha_1}\ldots\xi^{\alpha_s}\;\bar{\xi}^{\dot\beta_1}\ldots\bar{\xi}^{\dot\beta_s} \,,
\end{equation}
which satisfies also the equations \p{cond-10}:
\begin{equation}\label{cond-10d}
P^{2}\Psi =0\,,\qquad
\left(\frac{\partial}{\partial \bar\xi^{\dot\alpha}} P^{\dot\alpha\beta}\frac{\partial}{\partial \xi^{\beta}}+\mu\right)\Psi =0\,.
\end{equation}
In addition, due to the presence of $\delta$-function in the
expression (\ref{Phi-expan0d}), this field satisfies the equation
\begin{equation}
\left( \xi^{\alpha}P_{\alpha\dot\beta}\bar\xi^{\dot\beta}-\mu\right)\Psi =0\,.
\label{cond-10d-2}
\end{equation}

Note that all component fields  $\varphi_{\alpha_1\ldots
\alpha_s\dot\beta_1\ldots\dot\beta_s}$ in the expansion \p{exp0} are
independent up to solution of the equation of motion. But this is
not so for the field \p{Phi-expan0d} containing the
$\delta$-function.

The presence of $\delta$-function in (\ref{Phi-expan0d}) implies that
the equality $\xi p\bar\xi=\mu$ is fulfilled in this expression.
We pass on as before to the light-cone system (\ref{sys-st-m}). Then this equality takes the form
\begin{equation}
\xi^1\bar\xi^{\dot 1} =\frac{\mu}{2E}=\mathrm{const}\,.
\label{xi2}
\end{equation}
Consequently, the terms
\begin{equation} \label{1}
\xi^1\bar\xi^{\dot 1}\varphi_{1\dot1}+\frac{1}{2}\xi^1\xi^1\bar\xi^{\dot 1}\bar\xi^{\dot 1}\varphi_{11\dot1\dot1}+
\frac{1}{3!}\xi^1\xi^1\xi^1\bar\xi^{\dot 1}\bar\xi^{\dot 1}\bar\xi^{\dot 1}\varphi_{111\dot1\dot1\dot1}+\ldots
\end{equation}
are absorbed by the field $\varphi_0$ after its redefinition,
the terms
\begin{equation} \label{2}
\frac{1}{2}\xi^1\xi^1\bar\xi^{\dot 1}\bar\xi^{\dot 2}\varphi_{11\dot1\dot2}+
\frac{1}{3!}\xi^1\xi^1\xi^1\bar\xi^{\dot 1}\bar\xi^{\dot 1}\bar\xi^{\dot 2}\varphi_{111\dot1\dot1\dot2}+\ldots
\end{equation}
are absorbed by the term $\xi^1\bar\xi^{\dot 2}\varphi_{1\dot2}$,
the terms
\begin{equation} \label{3}
\frac{1}{2}\xi^1\xi^2\bar\xi^{\dot 1}\bar\xi^{\dot 1}\varphi_{12\dot1\dot1}+
\frac{1}{3!}\xi^1\xi^1\xi^2\bar\xi^{\dot 1}\bar\xi^{\dot 1}\bar\xi^{\dot 1}\varphi_{112\dot1\dot1\dot1}+\ldots
\end{equation}
are absorbed by the field $\xi^2\bar\xi^{\dot 1}\varphi_{2\dot1}$, etc.

Thus, if we leave independent fields in the expansion (\ref{Phi-expan0d}),
then the fields present in the expressions (\ref{1}), (\ref{2}), (\ref{3}) can be set equal to zero:
\begin{equation} \label{1a}
\!\!\!\!\!\!\! \varphi_{1\dot1}=\varphi_{11\dot1\dot1}=\varphi_{111\dot1\dot1\dot1}=\ldots=0\,,\quad
\varphi_{11\dot1\dot2}=\varphi_{111\dot1\dot1\dot2}=\ldots=0\,,\quad
\varphi_{12\dot1\dot1}=\varphi_{112\dot1\dot1\dot1}=\ldots=0\,,\ldots
\end{equation}
But these fields are precisely those fields in (\ref{exp0}) that led to the (infinite) multiplicativity of the spectrum.
As a result, in this case, with the $\delta$-function, the spectrum consists of states
\begin{equation}\label{exp01}
\underbrace{\varphi_0}_{0}\,,\qquad \underbrace{\varphi_{1\dot2}}_{1}\,,\underbrace{\varphi_{2\dot1}}_{-1}\,,\qquad
\underbrace{\varphi_{11\dot2\dot2}}_{2}\,,\underbrace{\varphi_{22\dot1\dot1}}_{-2}\,,\qquad \ldots
\end{equation}
Thus, in the spectrum, all helicities are present once and we get an irreducible representation of the infinite spin.

Note that the analysis of the second equation in (\ref{cond-10d}) (second equation in \p{cond-10})
is carried out in exactly the same way as in the case without the $\delta$-function.
In particular, for convenience, we can impose the conditions (\ref{1a}) in the solution presented in \p{cond-10-2a}.

\subsubsection*{B.1.3 \ Gauged field}

From the definition \p{Phi-expan0d} we get that the field
$\Psi(p,\xi,\bar{\xi})$ is not changed under the following
transformations of the field $\Phi(p,\xi,\bar{\xi})$:
\begin{equation} \label{Phi-trans}
\delta\Phi(p,\xi,\bar{\xi}) =
(\xi p\bar\xi-\mu)\,\Lambda(p,\xi,\bar{\xi}) \,,
\end{equation}
where the field $\Lambda(p,\xi,\bar{\xi})$ satisfies the equations \p{cond-10} or, the same, \p{cond-10d}.

We pass on as before to the light-cone system (\ref{sys-st-m}). Then
the field $\Phi(p,\xi,\bar{\xi})$ in \p{Phi-trans} is represented by
the formulae \p{cond-10-2a}, \p{cond-10-2b}, \p{cond-10-2c},
\p{cond-10-2e} due to the equation \p{cond-10d}. The field
$\Lambda(p,\xi,\bar{\xi})$ has the same expression with replacement
the fields $f^{(\pm k)}_l(E)$ by $\lambda^{(\pm k)}_l(E)$.

Due to the gauge transformation we can eliminate the field $f^{(\pm k)}_l(E)$, $l\geq 1$
by the gauge field $\lambda^{(\pm k)}_{l-1}(E)$, in full accordance with the conditions \p{1a}.
As result, residual physical fields are $f^{(\pm k)}_0(E)$, as in \p{exp01}.

Now we describe how the field towers $f^{(\pm k)}_l(E)$, $l\neq 0$,
preserving the first component fields $f^{(\pm k)}_0(E)$ in the
expansions are eliminated by gauge transformations. Let us
demonstrate it on the tower $f^{(0)}_l(E)$, $l=0,1,\ldots,\infty$,
which are grouped in the field
\begin{equation} \label{F0-exr}
F^{(0)}(E;\xi^{1},\bar\xi^{\dot1}) = f_0(E)+\rho f_1(E) +\rho^2 f_2(E) +\rho^3 f_3(E) + \cdots\,,
\end{equation}
where $\xi^{1}=\sqrt{\rho}\exp(i\theta)$ (we omit index $(0)$ in
$f^{(0)}_l(E)$). First terms of the gauge field are given by
expansion:
\begin{equation} \label{L0-exr}
\Lambda^{(0)}(E;\xi^{1},\bar\xi^{\dot1}) = \lambda_0(E)+\rho
\lambda_1(E) +\rho^2 \lambda_2(E) +\rho^3 \lambda_3(E) + \cdots\,.
\end{equation}
The transformation \p{Phi-trans} gives
\begin{equation} \label{f0-trans}
\delta f_0+\rho \delta f_1 +\rho^2 \delta f_2 +\rho^3 \delta f_3 + \cdots =
-\mu\lambda_0+\rho( 2E\lambda_0-\mu\lambda_1) +\rho^2 ( 2E\lambda_1-\mu\lambda_2) + \cdots\,.
\end{equation}

The quadratic integration of the fields $f_l(E)=f^{(0)}_l(E)$  with
respect to the integration measure $\displaystyle{\delta(p^-)dp^-\,
\delta(2E\rho-\nu)d\rho\,\frac{dE}{E}=\delta(p^-)dp^-
\,\delta(\rho-\frac{\nu}{2E})d\rho\,\frac{dE}{2E^2}}$ implies the
following asymptotic behavior of these fields in the lower energy
boundary
\begin{equation} \label{as-pr-f}
f_l(E)|_{E\to 0} = o(E^{l+1})\,.
\end{equation}
At following asymptotic behavior of gauge fields
\begin{equation} \label{as-pr-lambda}
\lambda_l(E)|_{E\to 0} = o(E^{l})\,,
\end{equation}
the transformations \p{f0-trans} kill all the fields $f_l(E)$,
$l\geq 1$, except the field $f_0(E)$.

Gauge removal of the fields $f^{(\pm k)}_l(E)$, $l\neq 0$ at $k\neq
0$ is performed in the similar way.

\subsection*{B.2 \ Half-integer spins}

The difference between the case of half-integer spins and the case
of integer spins is the presence of an external spinor index in the
generalized field, which is described by additional spinor $\zeta$.
As a result, in the case of the half-integer spin, the extraction of
an irreducible representation requires the use of the additional
conditions for the infinite spin field.

So, let us consider the field
\begin{equation}\label{Phi-expan12}
\tilde\Phi(p,\zeta,\xi,\bar{\xi}) :=\zeta^\gamma\Phi_\gamma(p,\xi,\bar{\xi}) =
\zeta^\gamma\sum_{s=0}^\infty
\frac{1}{s!}\;\varphi_{\gamma,\alpha_1\ldots \alpha_s\dot\beta_1\ldots\dot\beta_s}(p)\;
\xi^{\alpha_1}\ldots\xi^{\alpha_s}\;\bar{\xi}^{\dot\beta_1}\ldots\bar{\xi}^{\dot\beta_s} \,,
\end{equation}
which is subjected by the equations
\begin{equation}\label{cond-12}
P^{\dot\alpha\gamma}\frac{\partial}{\partial \zeta^{\gamma}}\tilde\Phi =0\,,\qquad
\left(\frac{\partial}{\partial \bar\xi^{\dot\alpha}} P^{\dot\alpha\beta}\frac{\partial}{\partial \xi^{\beta}}+\mu\right)\tilde\Phi =0\,.
\end{equation}
The component fields $\varphi_{\gamma,\alpha_1\ldots
\alpha_s\dot\beta_1\ldots\dot\beta_s}(p)$ in \p{Phi-expan12} do not
have definite symmetry property with respect the index $\gamma$ and
the indices $\alpha$-s.

As first step in the solution of the equation \p{cond-12} we extract
in \p{Phi-expan12} the fully symmetric component fields by the
following expansion:
\begin{eqnarray}\label{Phi-expan12-ex}
\tilde\Phi(p,\zeta,\xi,\bar{\xi}) &=& \hat\Phi(p,\zeta,\xi,\bar{\xi}) \ + \ \zeta^\gamma\xi_\gamma\, \hat{\mathcal X}(p,\xi,\bar{\xi})\,,\\ [7pt]
\hat\Phi(p,\zeta,\xi,\bar{\xi}) &:=&
\zeta^\gamma\sum_{s=0}^\infty
\frac{1}{s!}\;\phi_{(\gamma\alpha_1\ldots \alpha_s)(\dot\beta_1\ldots\dot\beta_s)}(p)\;
\xi^{\alpha_1}\ldots\xi^{\alpha_s}\;\bar{\xi}^{\dot\beta_1}\ldots\bar{\xi}^{\dot\beta_s} \,,
\label{Phi-expan12-ex1}\\ [7pt]
\hat{\mathcal X}(p,\xi,\bar{\xi}) &:=&
\sum_{s=0}^\infty
\frac{1}{s!}\;\psi_{(\alpha_1\ldots \alpha_s)(\dot\beta_1\ldots\dot\beta_s\dot\beta_{s+1})}(p)\;
\xi^{\alpha_1}\ldots\xi^{\alpha_s}\;\bar{\xi}^{\dot\beta_1}\ldots\bar{\xi}^{\dot\beta_{s+1}} \,,
\label{Phi-expan12-ex2}
\end{eqnarray}
i.e. in \p{Phi-expan12} the fields are written as follows
\begin{equation} \label{co-fie-exp}
\varphi_{\gamma,\alpha_1\ldots \alpha_s\dot\beta_1\ldots\dot\beta_s}(p)=\phi_{(\gamma\alpha_1\ldots \alpha_s)(\dot\beta_1\ldots\dot\beta_s)}(p)
\ + \ s\,\epsilon_{\gamma(\alpha_1}\psi_{\alpha_2\ldots \alpha_s)(\dot\beta_1\ldots\dot\beta_s)}(p)\,.
\end{equation}

Similarly to the solution of the integer spin equations \p{cond-10},
the equations \p{cond-12} take the form
\begin{equation}\label{C1+12}
p^{\dot\delta\gamma}\varphi_{\gamma,\alpha(s)\dot\beta(s)}=0\,,\qquad
s\,p^{\dot\beta_s\alpha_s}\varphi_{\gamma,\alpha(s)\dot\beta(s)}=-\mu\varphi_{\gamma,\alpha(s-1)\dot\beta(s-1)}\,.
\end{equation}
in terms of the component fields
$\varphi_{\gamma,\alpha(k)\dot\beta(s)}:=\varphi_{\gamma,\alpha_1\ldots \alpha_s\dot\beta_1\ldots\dot\beta_s}$.

First equations in (\ref{C1+12}) is the massless Dirac equation and
in the light-cone system (\ref{sys-st-m}) it leads to
$$
\varphi_{2,\alpha(s)\dot\beta(s)}=\phi_{(2\alpha_1\ldots \alpha_s)(\dot\beta_1\ldots\dot\beta_s)}(E)
+ s\,\epsilon_{2(\alpha_1}\psi_{\alpha_2\ldots \alpha_s)(\dot\beta_1\ldots\dot\beta_s)}(E)
=0\,,\quad s=0,1,2,\ldots
$$
These equations express the fields $\phi$ with at least one undotted index $2$ in terms of the  fields $\psi$.

The second equations in (\ref{C1+12}), which are also written in the light-cone system
(\ref{sys-st-m}) in the form
$$
\begin{array}{l}
2Es\,\phi_{(1\alpha_1\ldots \alpha_{s-1}2)(\dot\beta_1\ldots\dot\beta_{s-1}\dot2)}(E)
+ 2Es^2\,\epsilon_{1(\alpha_1}\psi_{\alpha_2\ldots \alpha_{s-1}2)(\dot\beta_1\ldots\dot\beta_{s-1}\dot2)}(E)
 \\ [6pt]
\qquad + \ \mu \, \phi_{(1\alpha_1\ldots \alpha_{s-1})(\dot\beta_1\ldots\dot\beta_{s-1})}(E)
+ \mu (s-1)\,\epsilon_{1(\alpha_1}\psi_{\alpha_2\ldots \alpha_{s-1})(\dot\beta_1\ldots\dot\beta_{s-1})}(E)
=0\,,\quad  s =1,2,\ldots\,,
\end{array}
$$
express the fields $\psi$ with at least one dotted index $\dot2$ in terms of other fields.

Thus, in the solution of the equations (\ref{C1+12}), the following
fields
\begin{equation}\label{indep-fields-12m}
\phi_{(\underbrace{\mbox{\footnotesize $11\ldots 1$}}_{s+1})(\dot\beta_1\ldots\dot\beta_s)}(E)\,,\qquad
\chi_{(\alpha_1\ldots \alpha_s)(\underbrace{\mbox{\footnotesize $\dot1\dot1\ldots \dot1$}}_{s+1})}(E)\,,\qquad
 s=0,1,2,\ldots
\end{equation}
are independent.

Now consider the field $\tilde\Psi$ defined by the following expression
\begin{equation}\label{Psi-expan12}
\tilde\Psi(p,\zeta,\xi,\bar{\xi}) := \delta(\xi p\bar\xi-\mu)\tilde\Phi(p,\zeta,\xi,\bar{\xi}) \,,
\end{equation}
where the field $\tilde\Phi$ is defined by \p{Phi-expan12},
\p{Phi-expan12-ex}. Due to the presence of $\delta$-function in the
expression (\ref{Phi-expan12}), this field satisfies the equation
\begin{equation}\label{cond-del-12}
\left( \xi^{\alpha}P_{\alpha\dot\beta}\bar\xi^{\dot\beta}-\mu\right)\tilde\Psi =0
\end{equation}
in addition to the equations \p{cond-12}.

Analysis of the solutions of the equation \p{cond-del-12} has
already been done in Subsection\,B.1.2. It was shown there that the
fields having pairs of indices $1\dot1$ are not independent. Thus,
the following fields
\begin{equation}\label{indep-fields-12}
\phi_{(\underbrace{\mbox{\footnotesize $11\ldots 1$}}_{s+1})(\underbrace{\mbox{\footnotesize $\dot2\dot2\ldots \dot2$}}_{s})}(E)\,,\qquad
\psi_{(\underbrace{\mbox{\footnotesize $22\ldots 2$}}_{s})(\underbrace{\mbox{\footnotesize $\dot1\dot1\ldots \dot1$}}_{s+1})}(E)\,,\qquad
 s=0,1,2,\ldots
\end{equation}
are independent ones among the fields \p{indep-fields-12m} of a
generalized field \p{Psi-expan12}. Helicities of these fields are
equal to
\begin{equation}\label{helicities-indep-fields-12}
\frac12 +s\,,\qquad
-\frac12 -s\,,\qquad
 s=0,1,2,\ldots
\end{equation}

Second term in \p{psi-half-g} is described by generalized field
\begin{equation}\label{Ipsilon-expan12}
\tilde{\Upsilon}(p,\bar\zeta,\xi,\bar{\xi}) := \delta(\xi p\bar\xi-\mu)\tilde{\mathcal X}(p,\bar\zeta,\xi,\bar{\xi}) \,,
\end{equation}
where
\begin{equation}\label{Chi-expan12}
\tilde{\mathcal X}(p,\zeta,\bar\xi,\bar{\xi}) :=\bar\zeta^{\dot\gamma}{\mathcal X}_{\dot\gamma}(p,\xi,\bar{\xi}) =
\bar\zeta^{\dot\gamma}\sum_{s=0}^\infty
\frac{1}{s!}\;\bar\chi_{\dot\gamma,\alpha_1\ldots \alpha_s\dot\beta_1\ldots\dot\beta_s}(p)\;
\xi^{\alpha_1}\ldots\xi^{\alpha_s}\;\bar{\xi}^{\dot\beta_1}\ldots\bar{\xi}^{\dot\beta_s} \,.
\end{equation}
The field \p{Ipsilon-expan12} is subjected by the equations
\begin{equation}\label{cond-12-con}
\frac{\partial}{\partial \bar\zeta^{\dot\gamma}} P^{\dot\gamma\alpha}\tilde{\Upsilon} =0\,,\qquad
\left(\frac{\partial}{\partial \bar\xi^{\dot\alpha}} P^{\dot\alpha\beta}\frac{\partial}{\partial \xi^{\beta}}+\mu\right)\tilde{\Upsilon} =0\,,
\qquad \left( \xi^{\alpha}P_{\alpha\dot\beta}\bar\xi^{\dot\beta}-\mu\right)\tilde{\Upsilon} =0\,.
\end{equation}
Performing a similar analysis, we find that this field describes the
helicities
\begin{equation}\label{helicities-indep-fields-12a}
-\frac12 -s\,,\qquad
\frac12 +s\,,\qquad
 s=0,1,2,\ldots
\end{equation}

Thus, the field \p{psi-half-g}, consisting two fields
\p{Psi-expan12}, \p{Ipsilon-expan12}, describes a reducible
representation of an infinite spin that contains each helicity
twice.

We emphasize that the condition \p{cond-del-12} can be taken into
account due to local symmetry, as in the case of integer helicities
considered in the Subsection\,B.1.3.

\section*{Appendix\,C: \ Irreducible massless infinite half-integer spin representation}
\def\theequation{C.\arabic{equation}}
\setcounter{equation}0

In the space-time description,  the irreducible infinite
half-integer spin representation is described by the field
\cite{BFI,BFI19a}
\begin{equation}
\label{wf-st-tw-12}
\Psi_\alpha(x;\xi,\bar\xi) \ = \
\int d^4 \pi \, e^{\displaystyle \,i \pi_{\beta}\bar\pi_{\dot\beta} x^{\dot\beta\beta}}\,\pi_\alpha\,
\Psi_{tw}^{(-1/2)} (\pi,\bar\pi;\xi,\bar\xi)\,,
\end{equation}
where the twistor field of the infinite half-integer spin particle is
\begin{eqnarray}
\label{wf-tw-hel-12}
\Psi_{tw}^{(-\frac12)}( \pi,\bar \pi;\xi,\bar\xi) &=&
\delta\left((\pi\xi)(\bar\xi\bar\pi)-\mu\right)\,
e^{\displaystyle -iq_0/p_0}\,
\hat\Psi_{tw}^{(-\frac12)} (\pi,\bar \pi;\xi,\bar\xi)\,,
\\ [6pt]
\nonumber
&& \hat\Psi_{tw}^{(-\frac12)} \ = \ \psi^{(-\frac12)}( \pi,\bar \pi) +
\sum\limits_{k=1}^{\infty} (\bar\xi\bar\pi)^{k}\,
\psi^{(-\frac12+k)}(\pi,\bar \pi) +
\sum\limits_{k=1}^{\infty} (\pi\xi)^{k}\,
\psi^{(-\frac12-k)}(\pi,\bar \pi)\,.
\end{eqnarray}
Here
\begin{equation}
\label{w0-p0}
\frac{q_0}{p_0} \ = \
\frac{\sqrt{\mu}\sum\limits_{\alpha=\dot\alpha}(\pi_{\alpha}\bar\xi_{\dot\alpha} + \xi_{\alpha}\bar\pi_{\dot\alpha})}
{\sum\limits_{\beta=\dot\beta}\pi_{\beta}\bar\pi_{\dot\beta}} \,.
\end{equation}
Twistor field \p{wf-tw-hel-12} satisfies the equations
\begin{equation}
\label{equ-wf-tw}
i\,\pi_{\alpha}\,\frac{\partial}{\partial\xi_{\alpha}}\,\Psi_{tw}^{(-\frac12)} =  \sqrt{\mu}\,\Psi_{tw}^{(-\frac12)} \,,\qquad
i\,\bar\pi_{\dot\alpha}\,\frac{\partial}{\partial\bar\xi_{\dot\alpha}}\,\Psi_{tw}^{(-\frac12)}
=  \sqrt{\mu}\,\Psi_{tw}^{(-\frac12)} \,,
\end{equation}
\begin{equation}
\label{equ-wf-tw-2}
\left(\pi_{\alpha}\frac{\partial}{\partial \pi_{\alpha}} \ -\
\bar \pi_{\dot\alpha}\frac{\partial}{\partial \bar \pi_{\dot \alpha}} \ + \
\xi_{\alpha}\frac{\partial}{\partial \xi_{\alpha}} \ -\
\bar\xi_{\dot\alpha}\frac{\partial}{\partial \bar \xi_{\dot \alpha}}\right)\,
\Psi_{tw}^{(-\frac12)} \ =\   - \,  \Psi_{tw}^{(-\frac12)}\,.
\end{equation}
Last equation \p{equ-wf-tw-2} is equivalent to the invariance
property of the twistor field
\begin{equation}
\label{equ-wf-tw-3}
\Psi_{tw}^{(-\frac12)} ( \mathrm{e}^{i\gamma}\pi_{\alpha},\mathrm{e}^{-i\gamma}\bar \pi_{\dot\alpha};
\mathrm{e}^{i\gamma}\xi_{\alpha},\mathrm{e}^{-i\gamma}\bar\xi_{\dot\alpha}) \ =\
\mathrm{e}^{-i\gamma}\Psi_{tw}^{(-\frac12)} ( \pi_{\alpha},\bar \pi_{\dot\alpha};\xi_{\alpha},\bar\xi_{\dot\alpha})\,.
\end{equation}

In \cite{BFI} it was shown that the field \p{wf-st-tw-12} is the
solution to the equations of motion \p{psi-eq0}, \p{psi-eq1},
\p{psi-eq2}, \p{psi-eq12}.

The field, defined by the expression \p{wf-st-tw-12}, automatically
satisfies the additional irreducibility condition
\begin{equation}
\label{add-equ-irr-12}
\frac{\partial}{\partial \xi_{\alpha}}\,\Psi_\alpha(x;\xi,\bar\xi) \ =\ 0\,.
\end{equation}
To prove the  equation \eqref{add-equ-irr-12} we consider
\begin{eqnarray}\nonumber
\frac{\partial}{\partial \xi_{\alpha}}\,\Psi_\alpha(x;\xi,\bar\xi) & =&
\int d^4 \pi \, e^{\displaystyle \,i \pi_{\beta}\bar\pi_{\dot\beta} x^{\dot\beta\beta}}\,\pi_\alpha\,\frac{\partial}{\partial \xi_{\alpha}}\,
\Psi_{tw}^{(-1/2)} (\pi,\bar\pi;\xi,\bar\xi)
\\
&=& -i \sqrt{\mu}\int d^4 \pi \, e^{\displaystyle \,i \pi_{\beta}\bar\pi_{\dot\beta} x^{\dot\beta\beta}}\,
\Psi_{tw}^{(-1/2)} (\pi,\bar\pi;\xi,\bar\xi) \,, \label{add-equ-irr-12a}
\end{eqnarray}
where the first equation in \p{equ-wf-tw} was used. But in the
integrand of \p{add-equ-irr-12a}, the field $\Psi_{tw}^{(-1/2)}
(\pi,\bar\pi;\xi,\bar\xi)$ has the property \p{equ-wf-tw-3}, while
the rest quantity $d^4 \pi \, e^{\displaystyle \,i
\pi_{\beta}\bar\pi_{\dot\beta} x^{\dot\beta\beta}}$ is invariant.
Therefore, the integral \p{add-equ-irr-12a} is equal to zero
identically, that proves the the equation \eqref{add-equ-irr-12}.

Note that the equation \eqref{add-equ-irr-12} zeroes the field
\p{Phi-expan12-ex2} in the expansion \p{Phi-expan12-ex}.

It should also be emphasized that in this paper we consider the
fields which are the power series in the additional spinor variable
$\xi$, while in the paper \cite{BFI} there was considered different
class of space-time fields (see Appendix\,B in \cite{BFI}).

\end{document}